\documentclass[12pt,onecolumn]{IEEEtran}
\usepackage{dsfont}
\usepackage{setspace}
\usepackage{clrscode}
\usepackage{pict2e}

\usepackage{amsmath}
\usepackage{amssymb}
\usepackage[dvips]{graphicx}
\usepackage{epsfig}
\usepackage{epstopdf}
\usepackage{algorithm}
\usepackage{algorithmic}
\usepackage{caption}
\usepackage{amsthm}
\usepackage{subcaption}
\usepackage{dsfont}

\usepackage[ps2pdf,
bookmarks=false,
bookmarksnumbered=false, 
bookmarksopen=false, 
colorlinks=false]{}

\makeatletter

\newcommand{\Rmnum}[1]{\expandafter\@slowromancap\romannumeral #1@}
\makeatother

\newcommand{\SINR}{\text{SINR}}

\newcommand{\g}{\text{g}}

\newcommand{\PC}{\text{P}_{\text{C}}}
\newcommand{\RC}{\text{R}_{\text{C}}}

\newcommand{\BE}{\text{BE}}

\newcommand{\LOS}{\text{LOS}}
\newcommand{\NLOS}{\text{NLOS}}
\DeclareMathOperator\erf{erf}

\newcommand{\figsize}{.7}

\newtheorem{Theorem1}{Theorem}

\begin{document}
%
\title{Coverage in Heterogeneous Downlink Millimeter Wave Cellular Networks}

\author{Esma Turgut and M. Cenk Gursoy
\thanks{The authors are with the Department of Electrical
Engineering and Computer Science, Syracuse University, Syracuse, NY, 13244
(e-mail: eturgut@syr.edu, mcgursoy@syr.edu).}
\thanks{The material in this paper will be presented in part at the IEEE Global Communications Conference (Globecom), Washington, DC, Dec. 2016 \cite{GC16}.}}

\maketitle
\begin{spacing}{1.7}
\begin{abstract}
In this paper, we provide an analytical framework to analyze heterogeneous downlink mmWave cellular networks consisting of $K$ tiers of randomly located base stations (BSs) where each tier operates in a mmWave frequency band. Signal-to-interference-plus-noise ratio (SINR) coverage probability is derived for the entire network using tools from stochastic geometry. The distinguishing features of mmWave communications such as directional beamforming and having different path loss laws for line-of-sight (LOS) and non-line-of-sight (NLOS) links are incorporated into the coverage analysis by assuming averaged biased-received power association and Nakagami fading. By using the noise-limited assumption for mmWave networks, a simpler expression requiring the computation of only one numerical integral for coverage probability is obtained. Also, effect of beamforming alignment errors on the coverage probability analysis is investigated to get insight on the performance in practical scenarios. Downlink rate coverage probability is derived as well to get more insights on the performance of the network. Moreover, effect of deploying low-power smaller cells and the impact of biasing factor on energy efficiency is analyzed. Finally, a hybrid cellular network operating in both mmWave and $\mu$Wave frequency bands is addressed.
\end{abstract}

%

\thispagestyle{empty}


\section{Introduction}
There has been an exponential growth in mobile data and traffic in recent years due to, e.g., ever increasing use of smart phones, portable devices, and data-hungry multimedia applications. Limited available spectrum in microwave ($\mu$Wave) bands does not seem to be capable of meeting this demand in the near future, motivating the move to new frequency bands. Therefore, the use of large-bandwidth at millimeter wave (mmWave) frequency bands, between 30 and 300 GHz, becomes a good candidate for fifth generation (5G) cellular networks and has attracted considerable attention recently \cite{Rappaport1} -- \cite{Wang}.

Despite the great potential of mmWave bands, they have been considered attractive only for short range-indoor communication due to increase in free-space path loss with increasing frequency, and poor penetration through solid materials such as concrete and brick. However, these high frequencies may also be used for outdoor communication over a transmission range of about 150-200 meters as demonstrated by recent channel measurements \cite{Rappaport1}, \cite{Roh}, \cite{Ghosh}, \cite{Wang}. Also, comparable coverage area and much higher data rates than $\mu$Wave networks can be achieved provided that the base station density is sufficiently high and highly directional antennas are used \cite{Bai2}. With the employment of directional antennas, mmWave cellular networks can be considered as noise-limited rather than interference-limited \cite{Andrews}, \cite{Singh}, \cite{Kulkarni}, \cite{Akdeniz}, \cite{Marco2}. Also, another key feature of mmWave cellular networks is expected to be heterogeneity to have higher data rates and expanded coverage \cite{Rangan}.

A general model for heterogeneous cellular networks is described as a combination of $K$ spatially and spectrally coexisting tiers which are distinguished by their transmit powers, spatial densities, blockage models \cite{Dhillon}, \cite{Andrews2}. For example, high-power and low-density large-cell base stations (BSs) may coexist with denser but lower power small-cell BSs. Small cell BSs can help the congested large-cell BSs by offloading some percentage of their user equipments (UEs), which results in a better quality of service per UE \cite{Elsawy}. Moreover, to provide more relief to the large-cell network, cell range expansion technique which is enabled through cell biasing for load balancing was considered e.g., in \cite{Andrews2}, \cite{Damnjanovic}, \cite{Marco3}.

Several recent studies have also addressed heterogeneous mmWave cellular networks. In \cite{Ghadikolaei}, authors consider two different types of heterogeneity in mmWave cellular networks: spectrum heterogeneity and deployment heterogeneity. In spectrum heterogeneity, mmWave UEs may use higher frequencies for data communication while the lower frequencies are exploited for control message exchange. Regarding deployment heterogeneity, two deployment scenarios are introduced. In the stand-alone scenario, all tiers will be operating in mmWave frequency bands, while in the integrated scenario, $\mu$Wave network coexists with mmWave networks. A similar hybrid cellular network scenario is considered in \cite{Singh} for characterizing uplink-downlink coverage and rate distribution of self-backhauled mmWave cellular networks, and in \cite{Elshaer} for the analysis of downlink-uplink decoupling. In both papers, mmWave small cells are opportunistically used and UEs are offloaded to the $\mu$Wave network when it is not possible to establish a mmWave connection. In \cite{Rebato}, a hybrid spectrum access scheme (where exclusive access is used at frequencies in the 20/30 GHz range while spectrum sharing is used at frequencies around 70 GHz) is considered to harvest the maximum benefit from emerging mmWave technologies. A more general mathematical framework to analyze the multi-tier mmWave cellular networks is provided in \cite{Marco2}. In \cite{Maamari}, benefits of BS cooperation in the downlink of a heterogeneous mmWave cellular system are analyzed. Contrary to the hybrid scenario, each tier is assumed to operate in a mmWave frequency band in both \cite{Marco2} and \cite{Maamari}. Similarly, in this paper we consider a cellular network operating exclusively with mmWave cells, while, as we demonstrate in Section \ref{subsec:hybrid}, an extension to a hybrid scenario can be addressed and a similar analytical framework can be employed by eliminating the unique properties of mmWave transmissions in the analysis of the $\mu$Wave tier.

Stochastic geometry has been identified as a powerful mathematical tool to analyze the system performance of mmWave cellular networks due to its tractability and accuracy. Therefore, in most of the recent studies on heterogeneous and/or mmWave cellular networks, spatial distribution of the BSs is assumed to follow a point process and the most commonly used distribution is the Poisson point process (PPP) due to its tractability and accuracy in approximating the actual cellular network topology \cite{Elsawy}, \cite{Elsawy2}. In \cite{Elsawy2}, authors provide a comprehensive tutorial on stochastic geometry based analysis for cellular networks. Additionally, a detailed overview of mathematical models and analytical techniques for mmWave cellular systems are provided in \cite{Andrews3}. Since the path loss and blockage models for mmWave communications are significantly different from $\mu$Wave communications, three different states, namely line-of-sight (LOS), non-line-of-sight (NLOS) and  outage states, are considered for mmWave frequencies \cite{Akdeniz}, \cite{Marco2}. For analytical tractability, equivalent LOS ball model was proposed in \cite{Bai2}. In \cite{Singh}, authors considered probabilistic LOS ball model, which is more flexible than the LOS ball model to capture the effect of different realistic settings. In \cite{Marco2}, probabilistic LOS ball model is generalized to a two-ball model, which is based on path loss intensity matching algorithm. Path loss intensity matching approach to estimate the parameters of the path loss distribution is also employed in \cite{Marco2}, \cite{Marco4}, \cite{Marco5}.

In this paper, employing the tools from stochastic geometry and incorporating the distinguishing features of mmWave communications, we study heterogeneous donwlink mmWave cellular networks. Our main contributions can be summarized as follows:
\begin{enumerate}
\item A general expression of SINR coverage probability is derived for $K$-tier heterogeneous mmWave cellular networks by considering different Nakagami fading parameters for LOS and NLOS components, and employing the $D$-ball approximation for blockage modeling. Key differences from the previous work on mmWave heterogeneous cellular networks (e.g., \cite{Marco2}) are the following: We incorporate small-scale fading in the analysis and also use the more general $D$-ball model (rather than the two-ball model) for blockage modeling. Also, different from \cite{Marco2}
which considers the noise-limited approximation at the beginning of the analysis, we first provide a detailed and general analysis including interference calculation for both LOS and NLOS components, characterize the SINR coverage probability, and then identify under which conditions the noise-limited approximation is valid/accurate via numerical results. Moreover, we investigate the effect of biasing on mmWave heterogeneous cellular networks.

\item A simple expression for coverage probability for noise-limited case is obtained, and also a closed-form expression for some special values of LOS and NLOS path loss exponents is provided.

\item Energy efficiency analysis is conducted for $K$-tier heterogeneous mmWave cellular networks. Different from previous works, effect of biasing factor on energy efficiency is investigated for the first time in the literature.

\item Moreover, we describe how the analysis can be adapted to determine the coverage in hybrid cellular network scenarios, involving a $\mu$Wave large cell and mmWave smaller cells. We provide interesting observations and comparisons between the performances in the all-mmWave and hybrid scenarios. In particular, we highlight the impact of increased interference in the hybrid cellular network.

\end{enumerate}

The rest of the paper is organized as follows. In Section II, system model is introduced. In Section III, the total SINR coverage probability of the network is derived initially considering perfect beam alignment, and then in the presence of beamsteering errors. In Section IV, we provide several extensions of the main analysis. In particular, rate coverage probability is determined in Section IV-A, and energy efficiency is analyzed in Section IV-B. In Section IV-C, analysis of a hybrid cellular network scenario is provided. In Section V, numerical results are presented to identify the impact of several system parameters on the performance metrics. Finally, conclusions and suggestions for future work are provided in Section VI. Several proofs are relegated to the Appendix.

\section{System Model} \label{sec:system_model}
In this section, a $K$-tier heterogeneous downlink mmWave cellular network is modeled where the BSs in the $k^{\text{th}}$ tier are distributed according to a homogeneous PPP $\Phi_k$ of density $\lambda_k$ on the Euclidean plane for $k=1,2,\ldots,K$. BSs in all tiers are assumed to be transmitting in a mmWave frequency band, and the BSs in the $k^{\text{th}}$ tier are distinguished by their transmit power $P_k$, biasing factor $B_k$, and blockage model parameters. The UEs are also spatially distributed according to an independent homogeneous PPP $\Phi_u$ of density $\lambda_u$. Without loss of generality, a typical UE is assumed to be located at the origin according to Slivnyak's theorem \cite{Baccelli}, and it is associated with the tier providing the maximum average biased-received power.

In this setting, we have the following assumptions regarding the system model of the $K$-tier heterogeneous downlink mmWave cellular network:

\textit{Assumption 1 (Directional beamforming):} Antenna arrays at the BSs of all tiers and UEs are assumed to perform directional beamforming where the main lobe is directed towards the dominant propagation path while smaller sidelobes direct energy in other directions. For tractability in the analysis and similar to  \cite{Bai2}, \cite{Singh}, \cite{Marco2}, \cite{Bai3}, \cite{Bai4}, \cite{Wildman}, antenna arrays are approximated by a sectored antenna model, in which the array gains are assumed to be constant $M$ for all angles in the main lobe and another smaller constant $m$ in the side lobes \cite{Hunter}. Initially, perfect beam alignment is assumed in between UE and its serving BS\footnote{Subsequently, beamsteering errors are also addressed.}, leading to an overall antenna gain of $MM$. In other words, maximum directivity gain can be achieved for the intended link by assuming that the serving BS and UE can adjust their antenna steering orientation using the estimated angles of arrivals. Also, beam direction of the interfering links is modeled as a uniform random variable on $[0,2\pi]$. Therefore, the effective antenna gain between an interfering BS and UE is a discrete random variable (RV) described by
\vspace{-0.2cm}
\begin{equation}
    G=\left\{
                \begin{array}{ll}
                  MM &\text{with prob.} \; p_{MM}=\left(\frac{\theta}{2\pi}\right)^2\\
                  Mm &\text{with prob.} \; p_{Mm}=2\frac{\theta}{2\pi}\frac{2\pi-\theta}{2\pi} \\
                  mm &\text{with prob.} \; p_{mm}=\left(\frac{2\pi-\theta}{2\pi}\right)^2,
                \end{array}
              \right. \label{eq:antennagains}
\end{equation}
where $\theta$ is the beam width of the main lobe, and $p_{G}$ is the probability of having an antenna gain of $G \in \{MM, Mm, mm\}$.

\textit{Assumption 2 (Path loss model and blockage modeling):} Link between a BS and a typical UE can be either a line-of-sight (LOS) or non-line-of-sight (NLOS) link. However, according to recent results on mmWave channel modeling, an additional outage state can also be included to represent link conditions. Therefore, a link can be in a LOS, NLOS or in an outage state \cite{Akdeniz}. In a LOS state, BS should be visible to UE, i.e., there is no blockage in the link. On the other hand, in a NLOS state, blockage occurs in the link, and if this blockage causes a very high path loss, an outage state occurs, i.e, no link is established between the BS and the UE.

Consider an arbitrary link of length $r$, and define the LOS probability function $p(r)$ as the probability that the link is LOS. Using field measurements and stochastic blockage models, $p(r)$ can be modeled as $e^{-\gamma r}$ where decay rate $\gamma$ depends on the building parameter and density \cite{Bai1}. For analytical tractability, LOS probability function $p(r)$ can be approximated by step functions. In this approach, the irregular geometry of the LOS region is replaced with its equivalent LOS ball model. Approximation by step functions provides tractable but also accurate results \cite{Marco5}, \cite{Ding}. Authors in both \cite{Marco5} and \cite{Ding} employ piece-wise LOS probability functions and multi-ball ball models. Futhermore, in \cite{Marco5}, comparisons of the intensity measures of empirical models (in London and Manchester) and 3GPP-based models with their 3-ball counterpart approximation models have been provided and good matching accuracy has been observed.

\begin{figure}[!h]
\centering
  \includegraphics[width=0.35\textwidth]{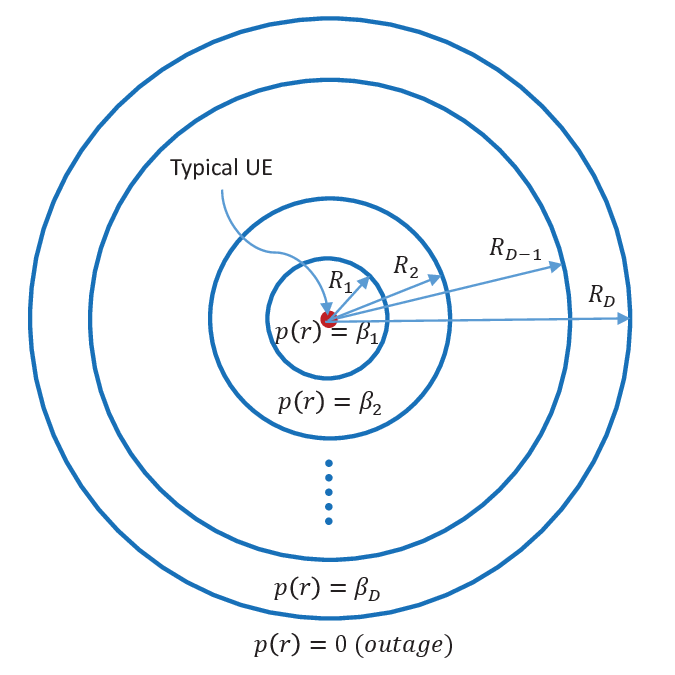}
  \caption{LOS ball model}
\label{D-ball}
\end{figure}

In this paper, we adopt a $D$-ball approximation model similar to the piece-wise LOS probability function approach proposed in \cite{Marco5}. As shown in Fig. \ref{D-ball}, a link is in LOS state with probability $p(r)=\beta_1$ inside the first ball with radius $R_1$, while NLOS state occurs with probability $1-\beta_1$. Similarly, LOS probability is equal to $p(r)=\beta_d$ for $r$ between $R_{d-1}$ and $R_d$ for $d=2,\ldots,D$, and all links with distances greater than $R_D$ are assumed to be in outage state.

Different path loss laws are applied to LOS and NLOS links. Thus, the path loss on each link in the $k$th tier can be expressed as follows:
\end{spacing}
\begin{equation}
    L_k(r)=\left\{
                \begin{array}{ll}
                 \begin{cases}
\kappa_1^L r^{\alpha_1^{k,L}}  \; \text{with prob. } \, \beta_{k1}
\\
\kappa_1^N r^{\alpha_1^{k,N}}  \; \text{with prob. } \, (1-\beta_{k1})
\end{cases} \vspace{0.1cm} \quad \text{if} \; r \le R_{k1}\\

                 \begin{cases}
\kappa_2^L r^{\alpha_2^{k,L}}  \; \text{with prob. } \, \beta_{k2}
\\
\kappa_2^N r^{\alpha_2^{k,N}}  \; \text{with prob. } \, (1-\beta_{k2})
\end{cases} \quad \text{if} \; R_{k1} \le r \le R_{k2} \\

\vdots \\
                 \begin{cases}
\kappa_D^L r^{\alpha_D^{k,L}}  \; \text{with prob. } \, \beta_{kD}
\\
\kappa_D^N r^{\alpha_D^{k,N}}  \; \text{with prob. } \, (1-\beta_{kD})
\end{cases} \quad \text{if} \; R_{k(D-1)} \le r \le R_{kD} \\

                 \text{outage} \quad \text{if} \; r \geq R_{kD},
                \end{array}
              \right.  \label{eq:Pathloss}
\end{equation}
\begin{spacing}{1.8}
\noindent where $\alpha_d^{k,L}$, and $\alpha_d^{k,N}$ are the LOS and NLOS path loss exponents for the $d$th ball of the $k$th tier, respectively, $\kappa_d^L$ and $\kappa_d^N$ are the path loss of LOS and NLOS links at a distance of 1 meter for the $d$th ball, respectively and $R_{kd}$ is the radius of the $d$th ball of the $k$th tier, for $d=1,\ldots,D$.

\subsection{Statistical Characterization of the Path Loss}
Let $\mathcal{N}_k=\{L_k(r)\}_{r \in \phi_k}$ denote the point process of the path loss between the typical UE and BSs in the $k$th tier. The characteristics of the typical UE which depend on the path loss can be determined by the distribution of $\mathcal{N}_k$ \cite{Blaszczyszyn}. Therefore, in Lemma 1 and Lemma 2 below, characterization of the complementary cumulative distribution function (CCDF) and the probability density function (PDF) of the path loss are provided.

\emph{Lemma 1:} The CCDF of the path loss from a typical UE to the BS in the $k$th tier can be formulated as
\begin{equation}
\bar{F}_{L_k}(x)=\mathbb{P}(L_k(r)>x)=\exp(-\Lambda_k([0,x))) \quad \text{for} \quad k=1,2,\ldots,K
\end{equation}
by applying the void probability theorem of PPPs \cite{Blaszczyszyn} with $\Lambda_k([0,x))$ defined as follows:
\begin{align}
\Lambda_k([0,x))&=\pi\lambda_k \sum_{d=1}^{D} \bigg(\beta_{kd}((R_{kd}^2-R_{k(d-1)}^2)\mathds{1}(x>\kappa_d^L R_{kd}^{\alpha_d^{k,L}})+((x/\kappa_d^L)^{\frac{2}{\alpha_d^{k,L}}}-R_{k(d-1)}^2) \nonumber \\
&\mathds{1}(\kappa_d^L R_{k(d-1)}^{\alpha_d^{k,L}}<x<\kappa_d^L R_{kd}^{\alpha_d^{k,L}}))+(1-\beta_{kd})((R_{kd}^2-R_{k(d-1)}^2)\mathds{1}(x>\kappa_d^N R_{kd}^{\alpha_d^{k,N}}) \nonumber \\
&+((x/\kappa_d^N)^{\frac{2}{\alpha_d^{k,N}}}-R_{k(d-1)}^2)\mathds{1}(\kappa_d^N R_{k(d-1)}^{\alpha_d^{k,N}}<x<\kappa_d^N R_{kd}^{\alpha_d^{k,N}}))\bigg), \label{intensity_function}
\end{align}
where $\mathds{1}(\cdot)$ is the indicator function and also note that $R_{k0}=0$.

\emph{Proof:} See Appendix \ref{Proof of Lemma 1}.

\emph{Lemma 2:} The CCDF of the path loss from the typical UE to the LOS/NLOS BS in the $k$th tier can be formulated as
\begin{equation}
\bar{F}_{L_{k,s}}(x)=\mathbb{P}(L_{k,s}(r)>x)=\exp(-\Lambda_{k,s}([0,x))) \quad \text{for} \quad k=1,2,\ldots,K
\end{equation}
where $s \in \{\LOS,\NLOS\}$ and $\Lambda_{k,s}([0,x))$ is defined for LOS and NLOS, respectively, as follows:
\begin{align}
\Lambda_{k,\LOS}([0,x))&=\pi\lambda_k \sum_{d=1}^{D} \bigg(\beta_{kd}((R_{kd}^2-R_{k(d-1)}^2)\mathds{1}(x>\kappa_d^L R_{kd}^{\alpha_d^{k,L}})+((x/\kappa_d^L)^{\frac{2}{\alpha_d^{k,L}}}-R_{k(d-1)}^2) \nonumber \\
&\mathds{1}(\kappa_d^L R_{k(d-1)}^{\alpha_d^{k,L}}<x<\kappa_d^L R_{kd}^{\alpha_d^{k,L}}))\bigg).
\label{intensity_function_LOS}
\end{align}
\begin{align}
\Lambda_{k,\NLOS}([0,x))&=\pi\lambda_k \sum_{d=1}^{D} \bigg((1-\beta_{kd})((R_{kd}^2-R_{k(d-1)}^2)\mathds{1}(x>\kappa_d^N R_{kd}^{\alpha_d^{k,N}}) \nonumber \\
&+((x/\kappa_d^N)^{\frac{2}{\alpha_d^{k,N}}}-R_{k(d-1)}^2)\mathds{1}(\kappa_d^N R_{k(d-1)}^{\alpha_d^{k,N}}<x<\kappa_d^N R_{kd}^{\alpha_d^{k,N}}))\bigg).
\label{intensity_function_NLOS}
\end{align}

\emph{Proof:} We can compute the intensities, $\Lambda_{k,\LOS}(\cdot)$ and $\Lambda_{k,\NLOS}(\cdot)$ of $\Phi_{k,\LOS}$ and $\Phi_{k,\NLOS}$, respectively, by following similar steps as in the proof of Lemma 1.\hfill $\square$

Also, the PDF of $L_{k,s}(r)$, denoted by $f_{k,s}$, which will be used in the following section is given by
\begin{equation}
\hspace{-0.05cm} f_{L_{k,s}}=-\frac{d\bar{F}_{L_{k,s}}(x)}{dx}=\Lambda_{k,s}^{\prime}([0,x)) \exp(-\Lambda_{k,s}([0,x)))
\end{equation}
where $\Lambda_{k,s}^{\prime}([0,x))$ is given as
\begin{equation}
    \Lambda_{k,s}^{\prime}([0,x))=\left\{
                \begin{array}{ll}
                  2\pi\lambda_k\sum_{d=1}^{D}\frac{ (x/\kappa_d^L)^{2/\alpha_d^{k,L}-1}}{\alpha_d^{k,L}} \big(\beta_{kd} \mathds{1}(\kappa_d^L R_{k(d-1)}^{\alpha_d^{k,L}}<x<\kappa_d^L R_{kd}^{\alpha_d^{k,L}})\big) \quad \text{for } s=\text{LOS} \\
                    2\pi\lambda_k\sum_{d=1}^{D}\frac{ (x/\kappa_d^N)^{2/\alpha_d^{k,N}-1}}{\alpha_d^{k,N}} \big((1-\beta_{kd}) \mathds{1}(\kappa_d^N R_{k(d-1)}^{\alpha_d^{k,N}}<x<\kappa_d^N R_{kd}^{\alpha_d^{k,N}})\big) \quad \text{for } s=\text{NLOS}
                \end{array}.
              \right. \label{derivativeof_intensity}
\end{equation}

The results of Lemma 1 and Lemma 2 are used in the calculation of association probabilities and SINR coverage probabilities in the following sections.

\subsection{Cell Association}
In this work, a flexible cell association scheme similarly as in \cite{Andrews2} is considered. In this scheme, UEs are assumed to be associated with the BS offering the strongest long-term averaged biased-received power. In other words, a typical UE is associated with a BS in tier-$k$ for $k=1,2,\ldots,K$ if
\begin{equation}
P_k G_k B_k L_k(r)^{-1} \geq P_j G_j B_j L_{min,j}(r)^{-1}, \text{for all} \; j=1,2,\ldots,K, j \neq k
\end{equation}
where $P$, $G$ and $B$ denote the transmission power, effective antenna gain of the intended link and biasing factor, respectively, in the corresponding tier (indicated by the index in the subscript), $L_k(r)$ is the path loss in the $k^{\text{th}}$ tier as formulated  in (\ref{eq:Pathloss}), and $L_{min,j}(r)$ is the minimum path loss of the typical UE from a BS in the $j$th tier. Antenna gain of the intended network $G$ is assumed to equal to $MM$ in all tiers for all-mmWave network, and it is equal to $M_{\mu}M$ for hybrid network where $M_{\mu}$ is defined as the antenna gain of the tier operating in $\mu$Wave frequency band. Although the analysis is done according to averaged biased-received power association, other association schemes like smallest path loss and highest average received power can be considered as well because they are special cases of biased association. When $B_k=1/(P_k G_k)$ for $k=1,2,\ldots,K$, biased association becomes the same as the smallest path loss association while $B_k=1$ for $k=1,2,\ldots,K$ corresponds to highest average received power association. In the following lemma, we provide the association probabilities with a BS in the $k$th tier using the result of Lemma 1.

\emph{Lemma 3:} The probability that a typical UE is associated with a LOS/NLOS BS in tier-$k$ for $k=1,2,\ldots,K$ is
\begin{equation}
\mathcal{A}_{k,s}= \int_0^{\infty} \Lambda_{k,s}^{\prime}([0,l_k)) e^{-\sum_{j=1}^{K} \Lambda_j\left([0,\frac{P_j G_j B_j}{P_k G_k B_k}l_k)\right)}dl_k  \qquad \text{for } s \in \{\text{LOS }, \text{NLOS}\} \label{Association_Prob}
\end{equation}
where $\Lambda_j([0,x))$, and $\Lambda_{k,s}^{\prime}([0,x))$ are given in (\ref{intensity_function}) and (\ref{derivativeof_intensity}), respectively.

\textit{Proof}: See Appendix \ref{Proof of Lemma 3}.

In the corollary below, we derive a closed-form expression for the association probability for a special case in order to provide several insights on the effects of different parameters on association probability.

\textbf{Corollary 1:} Consider a $2$-tier network with $1$-ball model for which the LOS probability is $\beta_{k1}=1$ and ball radius is $R_{k1}$ for tiers $k=1,2$. Further assume that $\alpha_{1}^{k,L}=2$ for $k=1,2$. Following several algebraic operations on (\ref{Association_Prob}), closed-form expressions for the probability that a typical UE is associated with a LOS BS in tier-$k$ for $k=1,2$, respectively, can be expressed as
\begin{small}
\begin{align}
&\hspace{-1cm}\mathcal{A}_{1,L}= \begin{cases}
      \frac{\lambda_1 P_1 G_1 B_1}{\sum_{j=1}^2 \lambda_j P_j G_j B_j}\big(1-e^{-\frac{\pi R_{11}^2}{P_1 G_1 B_1}\left(\sum_{j=1}^2 \lambda_j P_j G_j B_j\right)} \big), & \mbox{if } \frac{P_1 G_1 B_1}{P_2 G_2 B_2} R_{21}^2>R_{11}^2 \\
     \frac{\lambda_1 P_1 G_1 B_1}{\sum_{j=1}^2 \lambda_j P_j G_j B_j}\big(1-e^{-\frac{\pi R_{21}^2}{P_2 G_2 B_2}\left(\sum_{j=1}^2 \lambda_j P_j G_j B_j\right)} \big)+e^{-\frac{\pi R_{21}^2}{P_2 G_2 B_2}\sum_{j=1}^2 \lambda_j P_j G_j B_j }-e^{-\pi \sum_{j=1}^2 (\lambda_j R_{j1}^2) }, & \mbox{otherwise} \label{eq:closed-form-association1}
   \end{cases}
\end{align}
\vspace{-1cm}
\begin{align}
&\hspace{-1cm}\mathcal{A}_{2,L}= \begin{cases}
      \frac{\lambda_2 P_2 G_2 B_2}{\sum_{j=1}^2 \lambda_j P_j G_j B_j}\big(1-e^{-\frac{\pi R_{21}^2}{P_2 G_2 B_2}\left(\sum_{j=1}^2 \lambda_j P_j G_j B_j\right)} \big), & \mbox{if } \frac{P_2 G_2 B_2}{P_1 G_1 B_1} R_{11}^2>R_{21}^2 \\
     \frac{\lambda_2 P_2 G_2 B_2}{\sum_{j=1}^2 \lambda_j P_j G_j B_j}\big(1-e^{-\frac{\pi R_{11}^2}{P_1 G_1 B_1}\left(\sum_{j=1}^2 \lambda_j P_j G_j B_j\right)} \big)+e^{-\frac{\pi R_{11}^2}{P_1 G_1 B_1}\sum_{j=1}^2 \lambda_j P_j G_j B_j }-e^{-\pi \sum_{j=1}^2 \left(\lambda_j R_{j1}^2\right) }, & \mbox{otherwise}. \label{eq:closed-form-association2}
   \end{cases}
\end{align}
\end{small}
\normalsize

For sufficiently large values of $R_{11}$ and $R_{21}$, the terms involving the exponential functions in the above expressions decay to zero. Therefore, we can simplify (\ref{eq:closed-form-association1}) and (\ref{eq:closed-form-association2}) further and association probabilities can be approximated with the following expression (which also confirms the result in \cite{Andrews2}):
\begin{equation}
\mathcal{A}_{k,L} \approx \frac{\lambda_k P_k G_k B_k}{\sum_{j=1}^{K} \lambda_j P_j G_j B_j}. \label{eq:closed-form-association}
\end{equation}
Above in (\ref{eq:closed-form-association}), since the term $\sum_{j=1}^{K} \lambda_j P_j G_j B_j$ is a sum over all tiers and does not depend on $k$, a typical UE obviously prefers to connect to a tier with higher BS density, transmit power, effective antenna gain and biasing factor.


\section{SINR Coverage Analysis}
In this section, we develop a theoretical framework to analyze the downlink SINR coverage probability for a typical UE using stochastic geometry. Although an averaged biased-received power association scheme is considered for tier selection, the developed framework can also be applied to different tier association schemes.

\subsection{Signal-to-Interference-plus-Noise Ratio (SINR)}
The SINR experienced at a typical UE at a random distance $r$ from its associated BS in the $k$th tier can be written as
\begin{equation}
\SINR_k=\frac{P_k G_0 h_{k,0} L_k^{-1}(r)}{\sigma_k^2+\sum_{j=1}^K \sum_{i \in \Phi_{j}\setminus{B_{k,0}}} P_j G_{j,i} h_{j,i} L_{j,i}^{-1}(r)}
\end{equation}
where $G_0$ is the effective antenna gain of the link between the serving BS and UE which is assumed to be equal to $MM$, $h_{k,0}$ is the small-scale fading gain from the serving BS, $\sigma_k^2$ is the variance of the additive white Gaussian noise component. Interference has two components: intracell and intercell interference, where the first one is from the active BSs operating in the same cell with the serving BS, and the second one is from the BSs in other cells. A similar notation is used for interfering links, but note that the effective antenna gains $G_{j,i}$ are different for different interfering links as described in (\ref{eq:antennagains}). Since the small-scale fading in mmWave links is less severe than the conventional systems due to deployment of directional antennas, all links are assumed to be subject to independent Nakagami fading (i.e., small-scale fading gains have a gamma distribution). Parameters of Nakagami fading are $N_{\LOS}$ and $N_{\NLOS}$ for LOS and NLOS links, respectively, and they are assumed to be positive integers for simplicity. When $N_{\LOS}=N_{\NLOS}=1$, the Nakagami fading specializes to Rayleigh fading.
\vspace{-.2cm}

\subsection{SINR Coverage Probability}  \label{subsec:SINRcoverage} \label{sec:SINR_Coverage_Probability}
The SINR coverage probability $\PC^k(\Gamma_k)$ is defined as the probability that the received SINR is larger than a certain threshold $\Gamma_k>0$ when the typical UE is associated with a BS from the $k$th tier, i.e., $\PC^k(\Gamma_k)= \mathbb{P}(\SINR_k>\Gamma_k;t=k)$ where $t$ indicates the associated tier. Moreover, homogeneous PPP describing the spatial distribution of the BSs in each tier can be decomposed into two independent non-homogeneous PPPs: the LOS BS process $\Phi_{k,\LOS}$ and NLOS BS process $\Phi_{k,\NLOS}$. Therefore, the total SINR coverage probability $\PC$ of the network can be computed using the law of total probability as follows:
\begin{equation}
\PC=\sum_{k=1}^K \bigg[\PC^{k,\LOS}(\Gamma_k)\mathcal{A}_{k,\LOS}+\PC^{k,\NLOS}(\Gamma_k)\mathcal{A}_{k,\NLOS}\bigg], \label{CoverageProbability}
\end{equation}
where $s \in \{\LOS,\NLOS\}$, $\PC^{k,s}$ is the conditional coverage probability given that the UE is associated with a BS in $\Phi_{k,s}$,  and $\mathcal{A}_{k,s}$ is the association probability with a BS in $\Phi_{k,s}$, which is given in Lemma 3. In the next theorem, we provide the main result for the total network coverage.

\begin{Theorem1}: The total SINR coverage probability of the $K$-tier heterogeneous mmWave cellular network under Nakagami fading with parameter $N_s$ is
\begin{align}
\PC & \approx \sum_{k=1}^K \sum_{s \in \{\LOS,\NLOS\}} \int_0^{\infty} \sum_{n=1}^{N_s}(-1)^{n+1} {N_s \choose n} e^{-\frac{n \eta_s\Gamma_kl_{k,s}\sigma_k^2}{P_kG_0}} e^{-\sum_{j=1}^K \left(A+B+\Lambda_j\left( \left[0,\frac{P_j G_j B_j}{P_k G_k B_k} l_{k,s}\right) \right)\right)} \Lambda_{k,s}^{\prime}([0,l_{k,s}))dl_{k,s} \label{total_SINR_coverage}
\end{align}
where
\begin{equation}
A=\hspace{-0.8cm} \sum_{G \in \{MM,Mm,mm\}} \hspace{-0.5cm} p_{G} \int_{\frac{P_j B_j}{P_k B_k} l_{k,s}}^{\infty} \hspace{-0.5cm} \Psi\left(N_{\LOS},\frac{n \eta_{\LOS}\Gamma_kP_jGl_{k,s}}{P_kG_0tN_L}\right)\Lambda_{j,\LOS}(dt) \label{AA}
\end{equation}
and
\begin{equation}
B=\hspace{-0.8cm} \sum_{G \in \{MM,Mm,mm\}} \hspace{-0.5cm} p_{G}\int_{\frac{P_j B_j}{P_k B_k} l_{k,s}}^{\infty} \hspace{-0.5cm} \Psi\left(N_{\NLOS},\frac{n \eta_{\NLOS}\Gamma_kP_jGl_{k,s}}{P_kG_0tN_N}\right)\Lambda_{j,\NLOS}(dt) \label{BB}
\end{equation}
and $\Psi(N,x)=1-1/(1+x)^N$, $\eta_s=N_s(N_s!)^{-\frac{1}{N_s}}$, $ p_{G}$ is the probability of having antenna gain $G$ and is given in (\ref{eq:antennagains}).
\end{Theorem1}
\emph{Proof:} See Appendix \ref{Proof of Theorem 1}.

General sketch of the proof is as follows: First, SINR coverage probability is computed given that a UE is associated with a LOS/NLOS BS in the $k$th tier. Subsequently, each of the conditional probabilities are summed up to obtain the total coverage probability of the network. In determining the coverage probability given that a UE is associated with a LOS/NLOS BS in the $k$th tier, Laplace transforms of LOS/NLOS interferences from the $k$th tier are obtained using the thinning theorem and the moment generating function (MGF) of the gamma variable.

We also note that the result of Theorem 1 is an approximation due to the tail probability of a gamma
random variable. Although the characterization in Theorem 1 involves multiple integrals, computation can be carried out relatively easily by using numerical integration tools. Additionally, we can simplify the result further for the noise-limited case as demonstrated in the following corollaries, where computation of only a single integral is required in Corollary 2, and the result of Corollary 3 is in closed-form requiring only the computation of the $\erf$ function.

\subsection{Special Case: Noise-limited Network}
In the previous section, we analyzed the coverage probability for the general case in which both noise and interference are present. However, recent studies show that mmWave networks tend to be noise-limited rather than being interference-limited \cite{Andrews}, \cite{Singh}, \cite{Kulkarni}, \cite{Akdeniz}, \cite{Marco2}. Hence, in the following corollary coverage probability expression is provided assuming a noise-limited cellular network.

\textbf{Corollary 2:} When there is no interference, coverage probability of the network is given by
\begin{align}
\PC & \approx \sum_{k=1}^K \sum_{s \in \{\LOS,\NLOS\}} \int_0^{\infty} \sum_{n=1}^{N_s}(-1)^{n+1} {N_s \choose n} e^{-\frac{n \eta_s\Gamma_kl_{k,s}\sigma_k^2}{P_kG_0}} e^{-\sum_{j=1}^K (\Lambda_j([0,\frac{P_j G_j B_j}{P_k G_k B_k} l_{k,s})))} \Lambda_{k,s}^{\prime}([0,l_{k,s}))dl_{k,s}. \label{total_SINR_coverage_noise}
\end{align}
We obtain (\ref{total_SINR_coverage_noise}) directly from (\ref{total_SINR_coverage}) by making the terms $A$ and $B$, which arise from interference, equal to zero. Note that computation of (\ref{total_SINR_coverage_noise}) requires only a single integral.

\textbf{Corollary 3:} When $\alpha_d^{k,L}=2,\alpha_d^{k,N}=4$ $\forall k$ and $\forall d$, the SNR coverage probability of the network reduces to
\begin{align}
\PC &\approx\sum_{k=1}^K \left [ \PC^{k,\LOS}(\Gamma_k)\mathcal{A}_{k,\LOS}+\PC^{k,\NLOS}(\Gamma_k)\mathcal{A}_{k,\NLOS} \right ] \nonumber \\
&= \sum_{k=1}^K \sum_{n=1}^{N_{\LOS}}(-1)^{n+1} {N_{\LOS} \choose n} 2\pi\lambda_k \bigg [\sum_{n=1}^{N}\beta_{kn} \int_{\sqrt{\kappa_n^L}R_{k(n-1)}}^{\sqrt{\kappa_n^L}R_{kn}} x e^{-(a_L x^2+b_L x^2+c_L x+d_L)} dx \bigg ] \nonumber \\
&+ \sum_{k=1}^K \sum_{n=1}^{N_{\NLOS}}(-1)^{n+1} {N_{\NLOS} \choose n} \pi\lambda_k \bigg [\sum_{n=1}^{N}(1\!\!-\!\!\beta_{kn})\int_{\sqrt{\kappa_n^N}R_{k(n-1)}^2}^{\sqrt{\kappa_n^N}R_{kn}^2} \!\!\!\!\! e^{-(a_N x^2+b_N x^2+c_N x+d_N)} dx\bigg ]  \label{total_SINR_coverage_noise2}
\end{align}
where we define
\begin{align}
a_L&=\frac{n\eta_{\LOS}\Gamma_k\sigma_k^2}{P_k G_0}, \quad a_N=\frac{n\eta_{\NLOS}\Gamma_k\sigma_k^2}{P_k G_0} \nonumber \\
b_L&=b_N=\sum_{j=1}^K \pi \lambda_j \sum_{d=1}^{D}\beta_{jd} \mathds{1}(\zeta_d^L R_{j(d-1)}<x<\zeta_d^L R_{jd}) \nonumber \\
c_L&=c_N=\sum_{j=1}^K \pi \lambda_j \sum_{d=1}^{D}(1-\beta_{jd})\mathds{1}(\zeta_d^N R_{j(d-1)}^2<x<\zeta_d^N R_{jd}^2) \nonumber \\
d_L&=d_N=\sum_{j=1}^K \pi \lambda_j \sum_{d=1}^{D} ((R_{jd}^2-R_{j(d-1)}^2)(\beta_{jd} \mathds{1}(x>\zeta_d^L R_{jd})+(1-\beta_{jd})\mathds{1}(x>\zeta_d^N R_{jd}^2)) \nonumber \\
&-R_{j(d-1)}^2(\beta_{jd} \mathds{1}(\zeta_d^L R_{j(d-1)}<x<\zeta_d^L R_{jd})+(1-\beta_{jd})\mathds{1}(\zeta_d^N R_{j(d-1)}^2<x<\zeta_d^N R_{jd}^2))
\end{align}
where $\zeta_d^L=\sqrt{\kappa_d^L \frac{P_k G_k B_k}{P_j G_j B_j}}$ and $\zeta_d^N=\sqrt{\kappa_d^N \frac{P_kB_k}{P_jB_j}}$, and the indefinite integrals can computed as follows:
\begin{align}
 &\int x e^{-(a x^2+b x^2+cx+d)} dx = -\frac{e^{-x((a+b)x+c)-d}\bigg(\sqrt{\pi} c e^{\frac{(2(a+b)x+c)^2}{4(a+b)}} \text{erf}\big( \frac{2x(a+b)+c}{2\sqrt{a+b}} +2\sqrt{a+b}\big)\bigg)}{4(a+b)^{3/2}}
\end{align}
\begin{equation}
 \int e^{-(a x^2+b x^2+cx+d)} dx= -\frac{\sqrt{\pi} e^{\frac{c^2}{4(a+b)}-d} \text{erf}\big( \frac{2x(a+b)+c}{2\sqrt{a+b}}\big)}{2\sqrt{a+b}}.
\end{equation}
We obtain the coverage probability expression in (\ref{total_SINR_coverage_noise2}) by inserting $\alpha_d^{k,L}=2,\alpha_d^{k,N}=4$ $\forall k$ and $\forall d$ into (\ref{total_SINR_coverage_noise}) and applying a change of variables with $l_{k,\LOS}=l_{k,\NLOS}=x^2$.  Above, $\erf$ denotes the error function. Depending on the values of $\sqrt{\kappa_d^L}R_{k(d-1)}$, $\sqrt{\kappa_d^L}R_{kd}$, $\sqrt{\kappa_d^N}R_{k(d-1)}^2$ and $\sqrt{\kappa_d^N}R_{kd}^2$ for $k = 1,\ldots, K$ and $d = 1,\ldots, D$, values of $b_L$, $c_L$, $d_L$, $b_N$, $c_N$, and $d_N$ become either zero or some constant in the intervals of each integral. Hence, the given expression is practically in closed-form which requires only the computation of the error function $\erf(\cdot)$.

\subsection{SINR Coverage Probability Analysis In the Presence of Beamsteering Errors}
In Section \ref{sec:SINR_Coverage_Probability} and the preceding analysis, antenna arrays at the serving BS and the typical UE are assumed to be aligned perfectly and downlink SINR coverage probability is calculated in the absence of beamsteering errors. However, in practice, it may not be easy to have perfect alignment. Therefore, in this section, we investigate the effect of beamforming alignment errors on the coverage probability analysis. We employ an error model similar to that in \cite{Wildman}. Let $|\epsilon|$ be the random absolute beamsteering error of the transmitting node toward the receiving node with zero-mean and bounded absolute error $|\epsilon|_{\text{max}} \le \pi$. Due to symmetry in the gain $G_0$, it is appropriate to consider the absolute beamsteering error. The PDF of the effective antenna gain $G_0$ with alignment error can be explicitly written as \cite{Marco2}
\begin{align}
\hspace{-0.3cm} f_{G_0}(\g)& =F_{|\epsilon|}\left(\frac{\theta}{2}\right)^2\delta(\g-MM)+2F_{|\epsilon|}\left(\frac{\theta}{2}\right)\!\!\left(\!\!1-F_{|\epsilon|}\left(\frac{\theta}{2}\right)\!\!\right)  \delta(\g-Mm) +\left(\!\!1-F_{|\epsilon|}\left(\frac{\theta}{2}\right)\!\!\right)^2\delta(\g-mm),
\label{eq:PDFofG}
\end{align}
where $\delta(\cdot)$ is the Kronecker's delta function, $F_{|\epsilon|}(x)$ is the CDF of the misalignment error and (\ref{eq:PDFofG}) follows from the definition of CDF, i.e., $F_{|\epsilon|}(x)=\mathbb{P}\{|\epsilon|\le x\}$. Assume that the error $\epsilon$ is Gaussian distributed, and therefore the absolute error $|\epsilon|$ follows a half normal distribution with $F_{|\epsilon|}(x)=\text{erf}(x/(\sqrt{2}\sigma_{\BE}))$, where $\text{erf}(\cdot)$ again denotes the error function and $\sigma_{\BE}$ is the standard deviation of the Gaussian error $\epsilon$.

It is clear that total SINR coverage probability expression in (\ref{total_SINR_coverage}) depends on the effective antenna gain $G_0$ between the typical UE and the serving BS in each tier. Thus, total SINR coverage probability $\PC$ can be calculated by averaging over the distribution of $G_0$, $f_{G_0}(\g)$, as follows:
\begin{align}
\PC &= \int_0^{\infty}\PC(g)f_{G_0}(\g)d \g \nonumber \\
&=(F_{|\epsilon|}(\theta/2))^2 \PC(MM)+2(F_{|\epsilon|}(\theta/2))\bar{F}_{|\epsilon|}(\theta/2) \PC(Mm)+\bar{F}_{|\epsilon|}(\theta/2)^2 \PC(mm),
\end{align}
where we define $\bar{F}_{|\epsilon|}(\theta/2)=1-F_{|\epsilon|}(\theta/2)$.

\section{Extensions to Other Performance Metrics and Hybrid Scenario}
In this section, we provide extensions of our main analysis, and formulate other performance metrics using the SINR coverage probability expression obtained in the previous section to get more insights on the performance of the network. First, downlink rate coverage probability expression for a typical UE is obtained. Then, we formulate the energy efficiency metric. Finally, we address the hybrid scenario involving both $\mu$Wave and mmWave frequency bands.

\subsection{Rate Coverage Probability}
In this subsection, we derive the downlink rate coverage probability for a typical UE. Since rate characterizes the data bits received per second per UE, it is also an important performance metric like SINR as an indicator of the serving link quality, and it is one of the main reasons motivating the move to mmWave frequency bands \cite{Gupta}. Similar to SINR coverage probability, the rate coverage probability $\RC^k(\rho_k)$ is defined as the probability that the rate is larger than a certain threshold $\rho_k>0$ when the typical UE is associated with a BS from the $k$th tier. Therefore, the total rate coverage $\RC$ of the network can be computed as follows:
\begin{equation}
\RC=\sum_{k=1}^K \RC^{k}(\rho_k)\mathcal{A}_{k}, \label{RateCoverageProbability}
\end{equation}
where  $\mathcal{A}_{k}=\mathcal{A}_{k,L}+\mathcal{A}_{k,N}$ is the association probability with a BS in $\Phi_{k}$. Conditional rate coverage probability can be calculated in terms of SINR coverage probability as follows:
\begin{align}
\RC^{k}(\rho_k)=\mathbb{P}(\text{Rate}_k>\rho_k)&=\mathbb{P}\left(\frac{W}{N_k} \log(1+\SINR_k)>\rho_k\right) \nonumber \\
&=\mathbb{P}\left(\SINR_k>2^{\frac{\rho_k N_k}{W}}-1\right) \nonumber \\
&=\PC^k(2^{\frac{\rho_k N_k}{W}}-1)
\end{align}
where $\PC^k(\cdot)$ is the SINR coverage probability of the $k$th tier (analyzed in Section \ref{subsec:SINRcoverage}), the instantaneous rate of the typical UE is defined as $\text{Rate}_k=\frac{W}{N_k} \log(1+\SINR_k)$, and $N_k$, also referred to as load, denotes the total number of UEs served by the serving BS. Note that the total available resource $W$ at the BS is assumed to be shared equally among all UEs connected to that BS. Round-robin scheduling is the well known example of the schedulers resulting in such a fair partition of resources to each UE. The load $N_k$ can be found using the mean load approximation as follows \cite{Singh2}
\begin{equation}
N_k=1+\frac{1.28\lambda_u \mathcal{A}_k}{\lambda_k}.
\end{equation}

\subsection{Energy Efficiency Analysis}
The deployment of heterogeneous mmWave cellular networks consisting of multiple tiers with different sizes will provide an opportunity to avoid coverage holes and improve the throughput. Additionally, dense deployment of low-power small cells can also improve the energy efficiency of the network by providing higher throughput and consuming less power. Moreover, load biasing can increase the energy efficiency further by providing more relief to the large-cell BSs. With these motivations, we investigate the energy efficiency of the proposed heterogeneous network with $K$ tiers. First, we describe the power consumption model and area spectral efficiency for each tier, and then formulate the energy efficiency metric, using the SINR coverage probability expression derived in the previous section.
\subsubsection{Power Consumption Model}
Largest portion of the energy in cellular networks are consumed by BSs \cite{Hasan}. In practice, total BS power consumption has two components: the transmit power and static power consumption. Therefore, we can model the total power consumption per BS using linear approximation model as $P_{tot}=P_0+\Delta P$, where $1/\Delta$ is the efficiency of the power amplifier, and $P_0$ is the static power consumption due to signal processing, battery backup, site cooling etc., and $P$ corresponds to the transmit power \cite{Richter}. Using this model, average power consumption (per unit area) of BSs in the $k$th tier  can be expressed as
\begin{equation}\label{power_model}
  P_{avg,k}=\lambda_k (P_{0,k}+\Delta_k P_{k}).
\end{equation}
\subsubsection{Area Spectral Efficiency}
The area spectral efficiency (i.e., network throughput) can be defined as the product of the throughput at a given link and density of BSs, and for the $k$th tier it can be formulated as follows:
\begin{equation}\label{ASE}
\tau_k=\lambda_k \PC^k(\Gamma_k) \log_2(1+\Gamma_k),
\end{equation}
where $\PC^k(\Gamma_k)$ is the SINR coverage probability when the typical UE is associated with a BS from the $k$th tier. Also, note that we assume universal frequency reuse among all BSs from the each tier, meaning that BSs share the same bandwidth.
\subsubsection{Energy Efficiency Metric}
We can formulate the energy efficiency metric as the ratio of the total area spectral efficiency to the average network power consumption as follows:
\begin{equation}\label{EE}
  \text{EE}=\frac{\sum_{k=1}^K \tau_k}{\sum_{k=1}^K P_{avg,k}}=\frac{\sum_{k=1}^K \lambda_k \PC^k(\Gamma_k) \log_2(1+\Gamma_k)}{\sum_{k=1}^K \lambda_k (P_{0,k}+\Delta_k P_{k})} \quad \text{bps/Hz/W}
\end{equation}
where $ P_{avg,k}$ and $\tau_k$ are given in (\ref{power_model}) and (\ref{ASE}), respectively. Given the characterizations of the coverage probabilities in Section \ref{subsec:SINRcoverage}, energy efficiency can be computed easily as demonstrated with the numerical results in Section \ref{sec:num}.

\subsection{Analysis of Hybrid Cellular Network Scenario} \label{subsec:hybrid}
Although in the preceding analysis we consider a cellular network operating exclusively with mmWave cells, proposed analytical framework can also be employed in the analysis of a hybrid cellular network in which the large cell is operating in the lower $\mu$Wave frequency band, and smaller cells are operating in the mmWave frequency band. The reason for considering a hybrid scenario is that coexistence of mmWave cells with a traditional $\mu$Wave cellular network is a likely deployment scenario in the transition process to the cellular network operating exclusively with mmWave cells. This is especially so in the case of sparse deployment of cellular networks \cite{Singh}.    Considering this hybrid scenario, we have different antenna and path loss models in the large $\mu$Wave cell. Particulary, large-cell BSs employ also directional antennas also but with a smaller main lobe gain and larger beam width of the main lobe, i.e., we set $M_{\mu}=3$dB and $\theta=120^{\circ}$. Regarding the path loss model, all the links from the large-cell BSs to the UEs are assumed to be LOS links, i.e., there are no blockages between BSs and UEs. With these assumptions, the SINR coverage probability of the hybrid network is now given as
\begin{align}
\PC & \approx \sum_{s \in \{\LOS\}} \int_0^{\infty} \sum_{n=1}^{N_s}(-1)^{n+1} {N_s \choose n} e^{-\frac{n \eta_s\Gamma_1l_{1,s}\sigma_1^2}{P_1G_0}} e^{- \left(A(j=1)+B(j=1)\right)} e^{-\sum_{j=1}^K \left(\Lambda_j\left( \left[0,\frac{P_j G_j B_j}{P_1 G_1 B_1} l_{1,s}\right) \right)\right)} \Lambda_{1,s}^{\prime}([0,l_{1,s}))dl_{1,s}  \nonumber \\
 +&\sum_{k=2}^K \sum_{s \in \{\LOS,\NLOS\}} \int_0^{\infty} \sum_{n=1}^{N_s}(-1)^{n+1} {N_s \choose n} e^{-\frac{n \eta_s\Gamma_kl_{k,s}\sigma_k^2}{P_kG_0}}  e^{-\sum_{j=2}^K \left(A+B\right)} e^{-\sum_{j=1}^K \left(\Lambda_j\left( \left[0,\frac{P_j G_j B_j}{P_k G_k B_k} l_{k,s}\right) \right)\right)} \Lambda_{k,s}^{\prime}([0,l_{k,s}))dl_{k,s} \label{total_SINR_coverage_hybrid}
\end{align}
where the first term is the coverage probability of the large cell operating in $\mu$Wave frequency bands, the second term is the total coverage probability of smaller cells operating in mmWave frequency bands, and  $A$ and $B$ are given in (\ref{AA}) and (\ref{BB}), respectively. Note that since large cell and smaller cells are operating in different frequency bands, interference experienced in the large cell is only from other large-cell BSs in the same tier, and similarly interference in smaller cells is from only the BSs in the smaller cells.

\section{Simulation and Numerical Results} \label{sec:num}
In this section, we evaluate the theoretical expressions numerically. Simulation results are also provided to validate the the accuracy of the proposed model for the heterogeneous downlink mmWave cellular network as well as the accuracy of the analytical characterizations. In the numerical evaluations and simulations, unless otherwise stated, a 3-tier heterogeneous network is considered and the parameter values are listed in Table \ref{Table}. For this 3-tier scenario, $k=1$, $k=2$ and $k=3$ correspond to the microcell, picocell, and femtocell, respectively. In other words, a relatively high-power microcell network coexists with denser but lower-power picocells and femtocells. For the microcell network, $D$-ball approximation is used with $D=2$ and the ball parameters are rounded from the values presented in \cite{Marco2} for 28 GHz. For smaller cells, we also employ the two-ball approximation in which the inner ball only consists of LOS BSs, and in the outer ball, only NLOS BSs are present.

\begin{table}
\small
\caption{System Parameters}
\centering
  \begin{tabular}{| l | r|}
    \hline
    \textbf{Parameters} & \textbf{Values}  \\ \hline
    $\alpha_d^{k,L}$, $\alpha_d^{k,N}$ $\forall k$, $\forall d$ & 2, 4 \\ \hline
    $N_{\LOS}$, $N_{\NLOS}$& 3, 2 \\ \hline
    $M$, $m$, $\theta$ & 10dB, -10dB, $30^{\circ}$  \\ \hline
    $\lambda_1$, $\lambda_2$, $\lambda_3$, $\lambda_u$  & $10^{-5}$, $10^{-4}$, $5\times 10^{-4}$, $10^{-3}$ $(1/m^2)$ \\ \hline
    $P_1$, $P_2$, $P_3$ & 53dBm, 33dBm, 23dBm  \\ \hline
    $P_{0,1}$, $P_{0,2}$, $P_{0,3}$  & 130W, 10W, 5W \\ \hline
    $\Delta_1$, $\Delta_2$, $\Delta_3$  & 4, 6, 8 \\ \hline
    $B_1$, $B_2$, $B_3$ & 1, 1, 1 \\ \hline
    $[R_{11} R_{12}]$, $[\beta_{11} \beta_{12}]$ & [50 200], [0.8 0.2] \\ \hline
    $[R_{21} R_{22}]$, $[\beta_{21} \beta_{22}]$ & [40 60], [1 0] \\ \hline
    $[R_{31} R_{32}]$, $[\beta_{31} \beta_{32}]$ & [20 40], [1 0] \\ \hline
    $\Gamma_k$ $\forall k$ & 0dB \\ \hline
    $\text{Carrier frequency} (F_c)$ & 28 GHz  \\ \hline
    $\text{Bandwidth} (W)$ & 1GHz \\ \hline
    $\kappa_d^L=\kappa_d^N$ $\forall d$ & $(F_c/4\pi)^2$ \\ \hline
    $\sigma_k^2$ $\forall k$ & -174 dBm/Hz +10log10($W$) + 10 dB \\ \hline
  \end{tabular} \label{Table}
\end{table}

\begin{figure}
\centering
  \includegraphics[width=\figsize\textwidth]{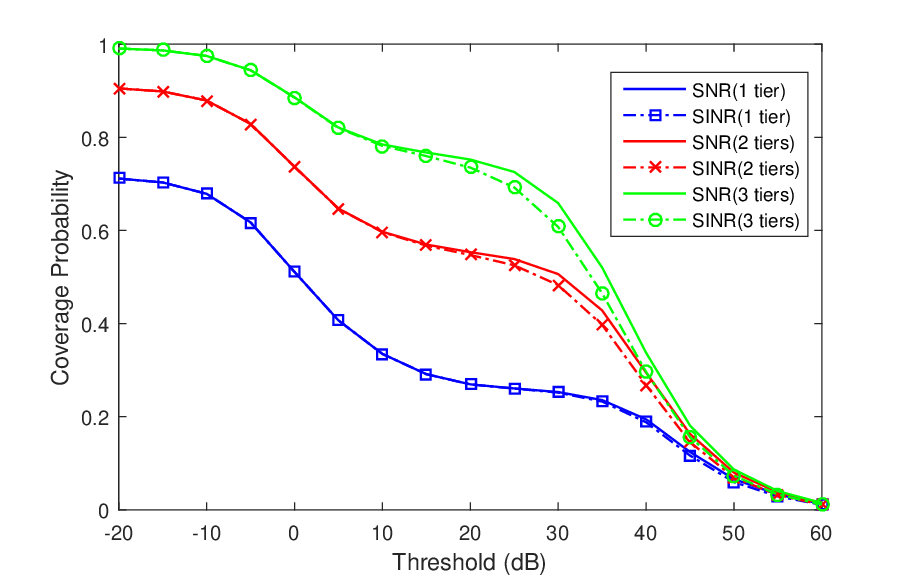}
  \caption{ Coverage Probability as a function of the threshold in dB comparison between SINR and SNR. \normalsize}
\label{Fig1}
\end{figure}

First, we investigate the noise-limited assumption of the mmWave cellular networks. In Fig. \ref{Fig1}, we plot the SINR and SNR coverage probabilities for three different number of tiers. When only microcell exists, since the interference is only from the same tier (i.e., microcell BSs), SINR and SNR coverage probabilities match with each other almost perfectly. As the number of tiers increases, the difference between SINR and SNR coverage probabilities become noticeable for higher values of the threshold because in a multi-tier scenario, interference is arising from BSs from different type of cells in different tiers as well. However, this performance gap is generally small and heterogeneous mmWave cellular networks can be assumed to be noise-limited (unless potentially the number of tiers is high). Also, note that as more tiers are added to the network, coverage probability increases significantly. Specifically, multi-tier network outperforms that with a single tier especially for small to medium values of the threshold.

\begin{figure}
\begin{subfigure}{.5\textwidth}
\centering
  \includegraphics[width=1\textwidth]{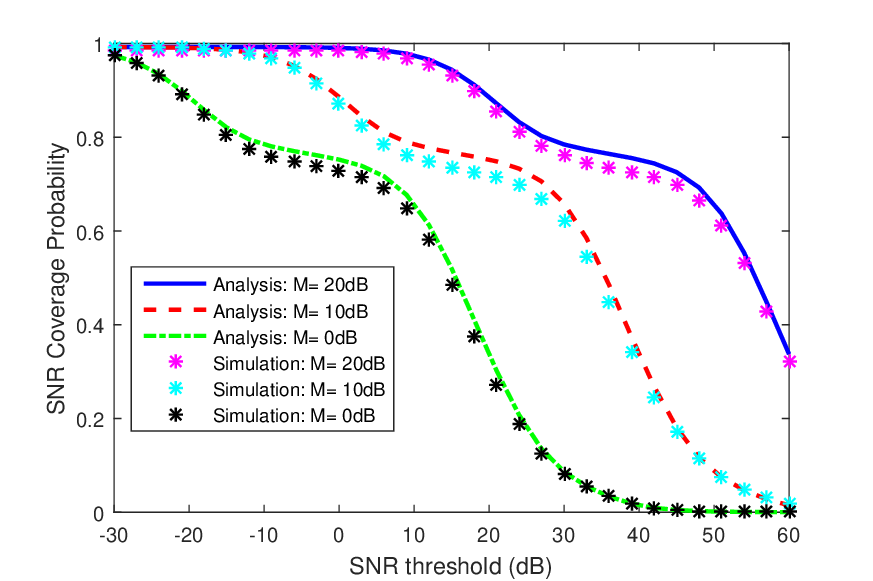}
\end{subfigure}
\begin{subfigure}{.5\textwidth}
\centering
  \includegraphics[width=1\textwidth]{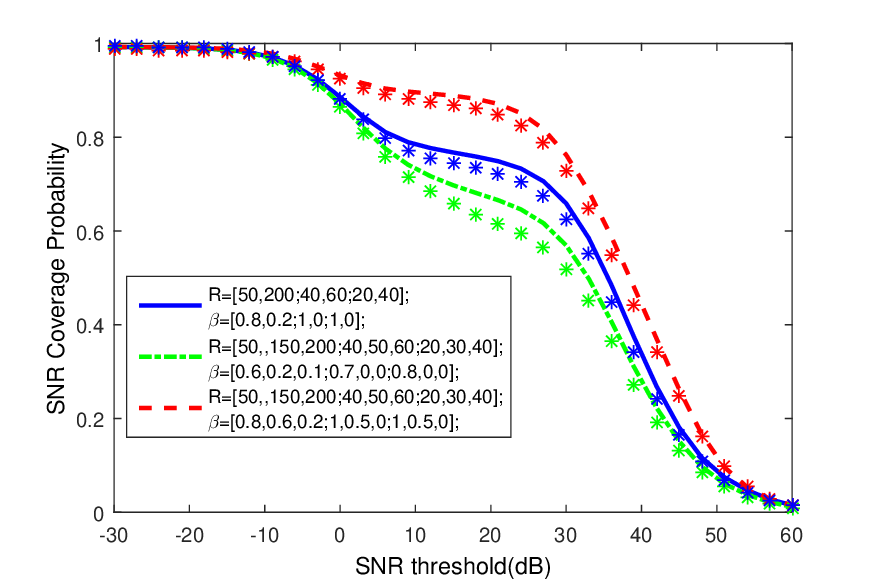}
\end{subfigure}
\caption{SNR Coverage Probability as a function of the threshold in dB for different values of (a) antenna main lobe gain $M$, (b) $D$-ball model parameters $R$ and $\beta$ \normalsize}
\label{Fig2}
\end{figure}

Since in Fig. \ref{Fig1} we show that the difference between SINR and SNR coverage probabilities are negligible even in multi-tier network scenarios, we henceforth consider the SNR coverage probabilities in the remaining simulation and numerical results. Next, we compare the SNR coverage probabilities for different values of the antenna main lobe gain $M$. As expected, better SNR coverage is achieved with increasing main lobe gain as shown in Fig. \ref{Fig2}(a). In Fig. \ref{Fig2}(b), SNR coverage probability is plotted for different parameters of the $D$-ball model. Solid line corresponds to the coverage probability with the default parameters, i.e. $2$-ball model with ball radii $(R_{11}, R_{12}), (R_{21}, R_{22}), (R_{31}, R_{32})$ in three tiers, respectively, and the corresponding $\beta$ parameters given as listed in Table \ref{Table} (and also provided in the legend of Fig. \ref{Fig2}(b)). Dashed line and dot-dashed lines are the coverage probabilities for the $3$-ball model with ball radii $(R_{11} = 50m, R_{12}=150m, R_{13}=200m), (R_{21}=40m, R_{22}=50m, R_{23}=60m), (R_{31}=20m, R_{32}=30m, R_{33}=40m)$ for the three tiers, respectively,  but with different LOS probabilities (denoted by $\beta$) as listed in the legend of Fig. \ref{Fig2}(b). Note that the LOS probabilities are higher for the case described by the dashed line (which implies that the signals are less likely to be blocked, for instance, as in a scenario with a less crowded environment and less buildings/blockages). Correspondingly, this high-LOS-probability 3-tier 3-ball model results in higher coverage probabilities. In the case of the dot-dashed curve, LOS probabilities are even smaller than those in the 2-ball model, resulting in degradation in the coverage probability. These numerical (and the accompanying simulation) results demonstrate that system parameters such as ball number and radii, and LOS probabilities have impact on the performance. Hence, appropriate modeling of the physical environment is critical in predicting the performance levels. Also note that, in Figs. \ref{Fig1}, \ref{Fig2}(a) and \ref{Fig2}(b), there are break points at certain points of the curves after which coverage probability degrades faster. In Fig. \ref{Fig2}(a), for  example, break points occur at approximately $70\%$ of the SNR coverage probability. These break points are occurring due to the assumption of the $D$-ball model. Finally, we also observe that simulation results very closely match the analytical results.

\begin{figure}
\centering
  \includegraphics[width=\figsize\textwidth]{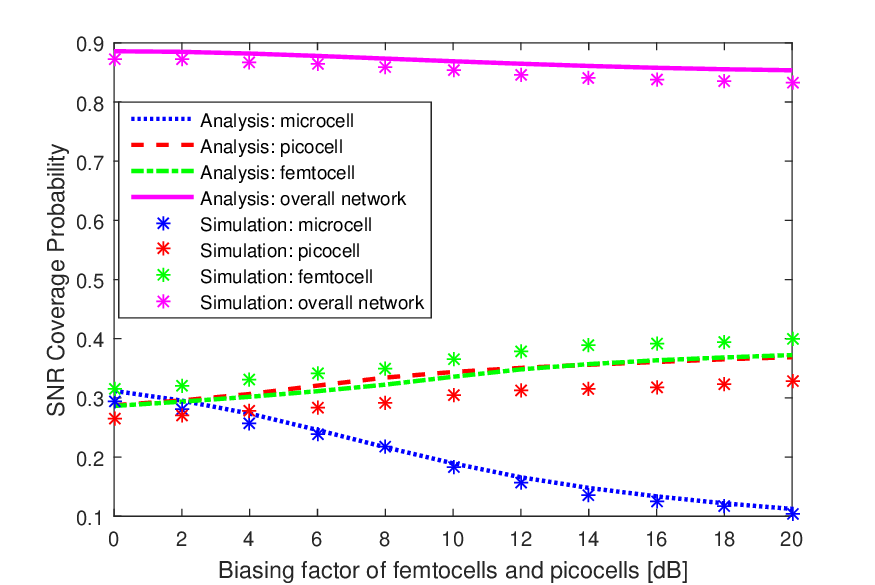}
  \caption{ SNR Coverage Probability as a function of the biasing factor of picocells and femtocells in dB ($B_1=0dB$).  \normalsize}
\label{Fig3}
\end{figure}

In Fig. \ref{Fig3}, we analyze the effect of biasing factor on the SNR coverage performance. We use the same biasing factor for picocells and femtocells, and no biasing for microcells. As the biasing factor increases, number of UEs associated with smaller cells increases resulting in an increase in coverage probabilities for picocells and femtocells while causing a degradation in the coverage performance of the microcell. This result is quite intuitive because with positive biasing, more UEs are encouraged to connect with the smaller cells. On the other hand, with biasing, UEs are associated with the BS not offering the strongest average received power, and thus the overall network coverage probability slightly decreases with the increasing biasing factor.

\begin{figure}
\centering
  \includegraphics[width=\figsize\textwidth]{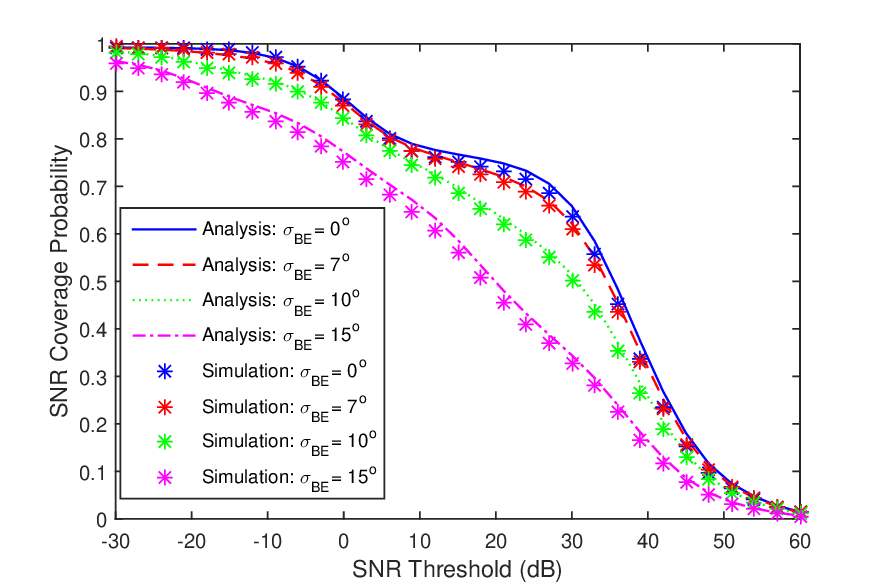}
  \caption{SNR Coverage Probability as a function of the threshold in dB for different alignment errors $\sigma_{BE}$. \normalsize}
\label{Fig4}
\end{figure}

In Fig. \ref{Fig4}, we show the effect of beam steering errors between the serving BS and the typical UE on the SNR coverage probability. As shown in the figure, coverage probability diminishes with the increase in alignment error standard deviation, and this deterioration becomes evident after $\sigma_{BE}=7^{\circ}$.

\begin{figure}
\centering
  \includegraphics[width=\figsize\textwidth]{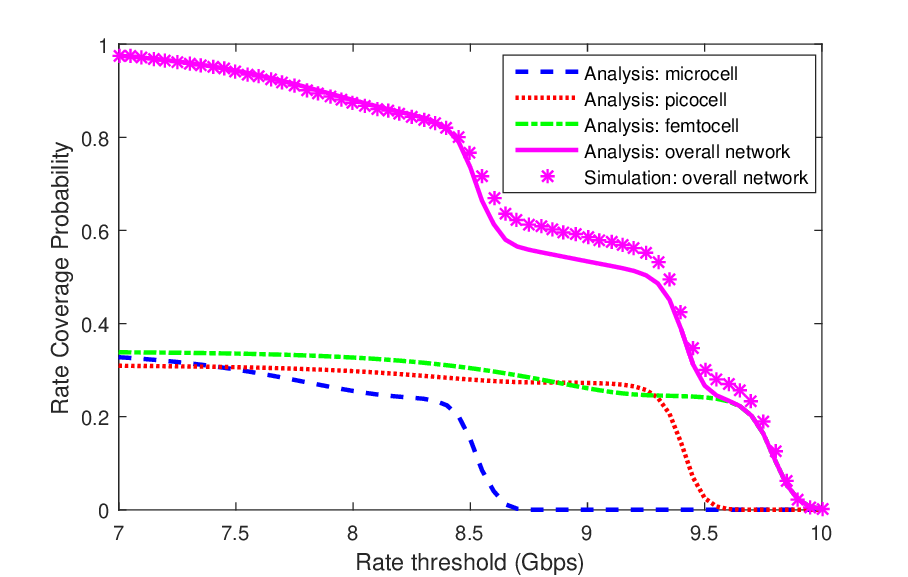}
  \caption{Rate Coverage Probability as a function of the threshold in Gbps.  \normalsize}
\label{Fig5}
\end{figure}

Fig. \ref{Fig5} shows the rate coverage probability as a function of the rate threshold. Rate coverage probability decreases with increasing rate threshold. Although there is a decrease in rate coverage probability, approximately $\%50$ percent coverage is provided for a rate of 9 Gbps, and 9.5 Gbps rate can be achieved with around $\%25$ percent coverage probability. Also, there are two transition lines in the overall network's rate coverage probability curve between 8.7-9.3 Gbps and 9.5-9.7 Gbps, respectively. The transition regions mainly distinguish the different tiers from each other. In other words, in the first transition region, microcell could not provide any rate coverage, and similarly picocells drop in the second region. Therefore, only femtocells can provide a rate greater than 9.5 Gbps.

\begin{figure}
\centering
  \includegraphics[width=\figsize\textwidth]{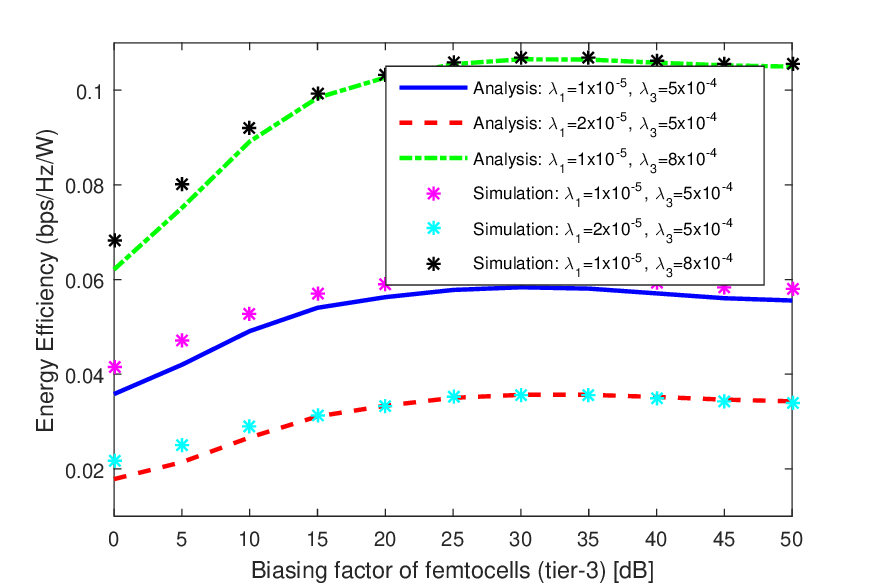}
  \caption{Energy Efficiency as a function of the biasing factor of femtocells in dB ($B_1=B_2=0dB$).   \normalsize}
\label{Fig6}
\end{figure}

In Fig. \ref{Fig6}, energy efficiency of a 3-tier heterogeneous downlink mmWave cellular network is plotted as a function of the biasing factor of femtocells for different values of the microcell and femtocell BS densities. As biasing factor increases, energy efficiency first increases and reaches its maximum point, and then it starts decreasing. Since biasing provides more relief to the high-power microcell and picocell BSs, energy efficiency initially improves with the increasing biasing factor due to the reduction in the total power consumption. However, further increase in the biasing factor causes a degradation in energy efficiency because the reduction in the total power consumption cannot compensate the decrease in the total coverage probability. Solid line corresponds to the energy efficiency curve for the default values of microcell and femtocell BS densities (given in Table \ref{Table}). When we increase the microcell BS density, energy efficiency degrades. On the other hand, when femtocell BS density is increased, energy efficiency improves. The reason is that introducing more low-power femto BSs is more energy efficient than the addition of more high-power micro BSs.

\begin{figure}
\begin{subfigure}{.5\textwidth}
\centering
  \includegraphics[width=1\textwidth]{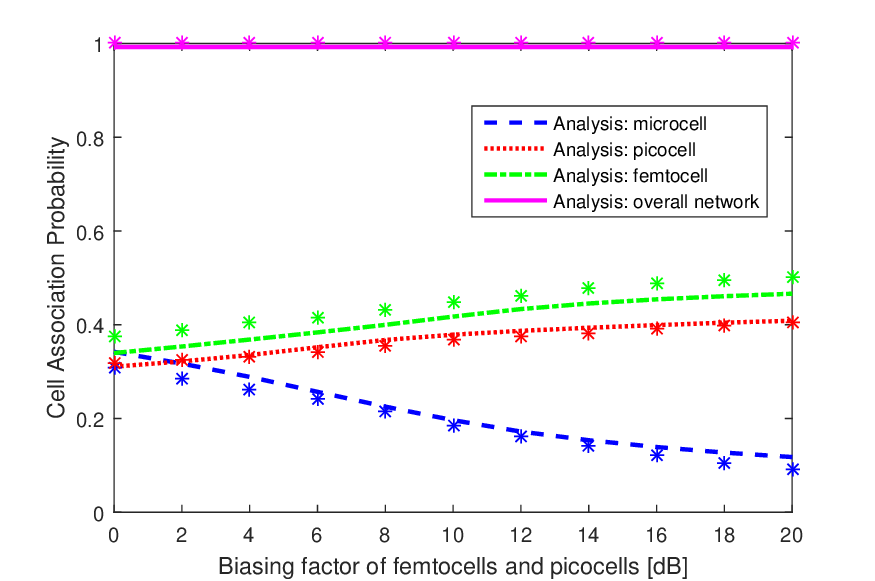}
\end{subfigure}
\begin{subfigure}{.5\textwidth}
\centering
  \includegraphics[width=1\textwidth]{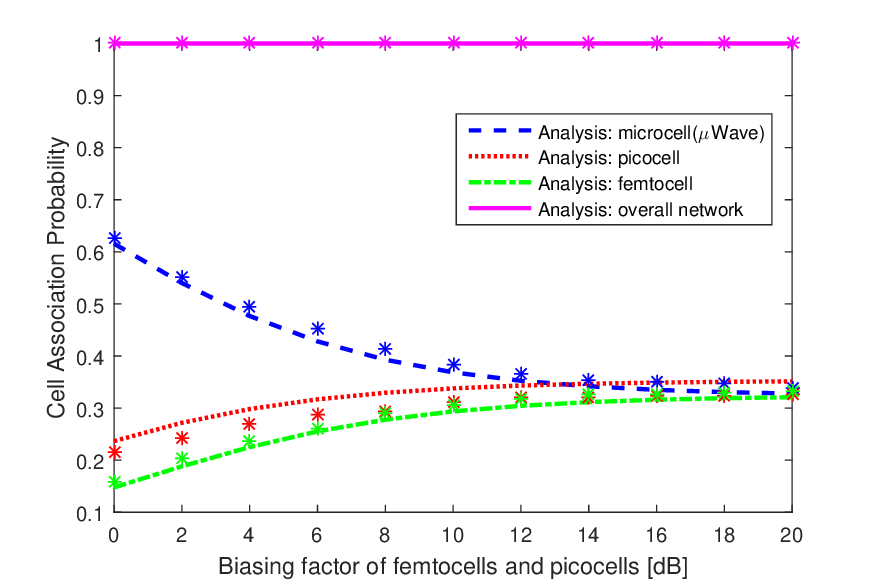}
\end{subfigure}
\caption{ Cell Association Probability for (a) all-mmWave network, (b) hybrid network as a function of the biasing factor of picocells and femtocells in dB ($B_1=0$dB).  \normalsize}
\label{Fig7}
\end{figure}

We plot the cell association probability for all-mmWave and hybrid network scenarios as a function of the biasing factor of picocells and femtocells in Fig. \ref{Fig7}(a) and Fig. \ref{Fig7}(b), respectively. In the hybrid network setup, we use the same parameters given in Table \ref{Table} with some differences for the microcell network operating at lower $\mu$Wave frequencies. More specifically, different from the previous figures, microcell BSs employ directional antennas with smaller main lobe gain, i.e., $M_{\mu}=3$dB and larger beam width $\theta=120^{\circ}$, and the links from these BSs to the UEs are assumed to be LOS links with $R_{11}=1500m$. Also, carrier frequency of the microcell network is $F_c=2$GHz and noise power is equal to $\sigma_1^2=-174 \text{ dBm/Hz} +10\log_{10}W + 10 \text{dB}$ where $W=20$MHz. Cell association probability of both all-mmWave and hybrid networks exhibit similar trends with the increase in biasing factor. However, association probability with microcell BSs (using $\mu$Wave frequencies) in the hybrid network is greater than that in the all-mmWave network despite the smaller antenna main lobe gain. Since average received power cell association criteria is employed for cell selection and microcell $\mu$Wave BSs have a larger LOS ball radius than smaller cells in the hybrid network, UEs tend to connect to $\mu$Wave BSs rather than mmWave BSs.

\begin{figure}
\centering
  \includegraphics[width=\figsize\textwidth]{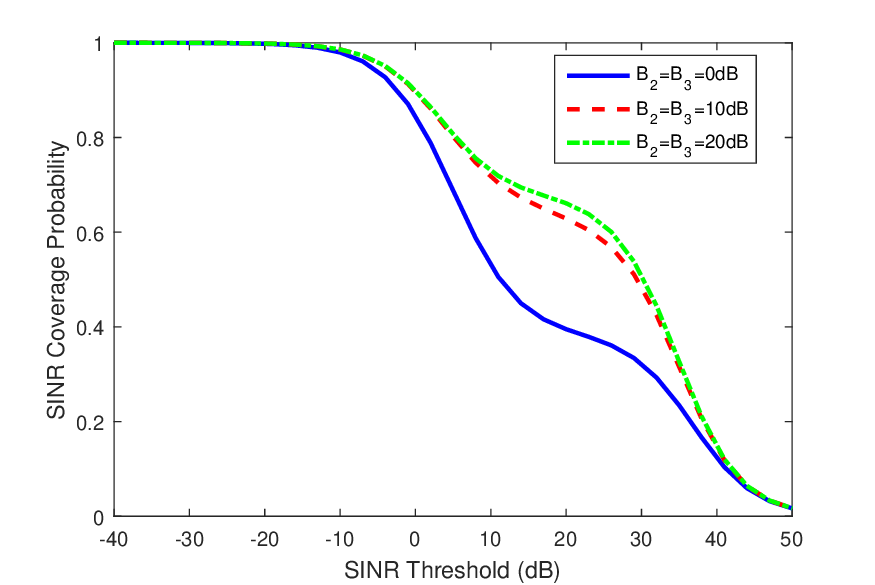}
  \caption{SINR Coverage Probability as a function of the threshold in dB for hybrid network for different biasing factor of picocells and femtocells in dB ($B_1=0dB$).  \normalsize}
\label{Fig8}
\end{figure}

In Fig. \ref{Fig8}, we plot the SINR coverage probability for hybrid network scenario as a function of the SINR threshold for different biasing factors of smaller cells. Although $\mu$Wave BSs provide higher average received power, overall SINR coverage probability becomes less when compared with the all-mmWave network scenario (as noticed when the coverage curves in Fig. \ref{Fig8} are compared with previous numerical results) because of the following reasons. Essentially, interference becomes an important concern with more impact in $\mu$Wave frequency bands, limiting the coverage performance. For instance, employment of omnidirectional antennas in microcell BSs is a critical factor (leading to increased interference and causing a poor coverage performance), along with having potentially more interfering $\mu$Wave microcell BSs due to longer possible link distances with LOS. Therefore, as noted before, overall SINR coverage probability is less than that in the all-mmWave network scenario. Also, as seen in the figure, SINR coverage probability increases as the biasing factor for the picocells and femtocells are increased (contrary to the previous observations in the all-mmWave network scenario where an increase in the biasing factor of the picocells and femtocells has slightly reduced the overall network coverage probability as seen in Fig. \ref{Fig3}). This again verifies the reasoning provided above. Specifically, with larger biasing factors, more UEs connect to the picocells and femtocells operating in the mmWave bands, and experience improved coverage due to employment of directional antennas and noise-limited nature of mmWave cells.

\begin{figure}
\centering
  \includegraphics[width=\figsize\textwidth]{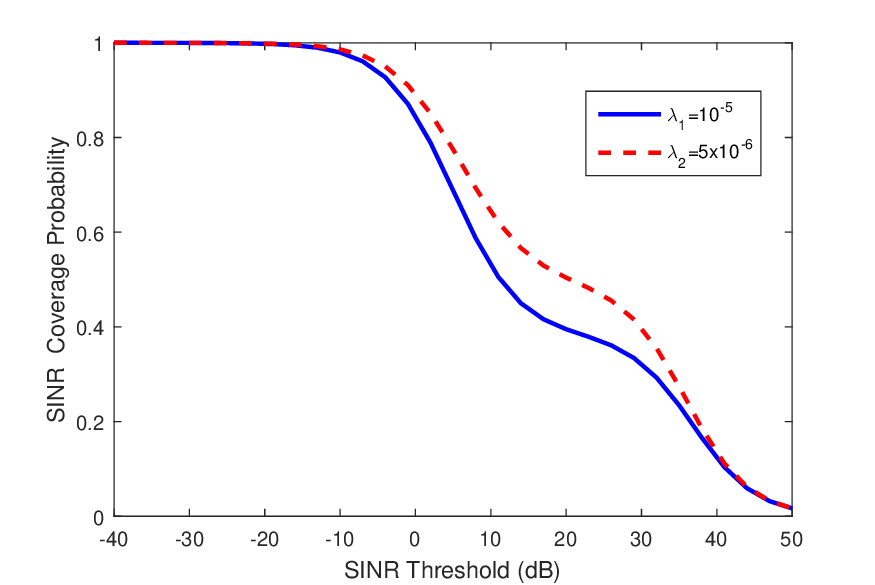}
  \caption{SINR Coverage Probability as a function of the threshold in dB for hybrid network for different density of microcells $\lambda_1$.  \normalsize}
\label{Fig9}
\end{figure}

Fig. \ref{Fig9} shows the effect of microcell BS density on the SINR coverage performance again for the hybrid scenario. Same parameter values are used as in Fig. \ref{Fig8} but with no biasing. We notice in this figure that coverage probability increases with decreasing microcell BS density due to the fact that when there is a smaller number of microcell BSs, interference from other BSs transmitting at the $\mu$Wave frequency band decreases.

\section{Conclusion}
In this paper, we have provided a general analytical framework to compute the SINR and rate coverage probabilities  in heterogeneous downlink mmWave cellular networks composed of $K$ tiers. Moreover, we have studied the energy efficiency metric and analyzed the effect of biasing on energy efficiency. Directional beamforming with sectored antenna model and $D$-ball approximation for blockage model have been considered in the analysis. BSs of each tier and UEs are assumed to be distributed according to independent PPPs, and UEs are assumed to be connected to the tier providing the maximum average biased-received power. Numerical results show that mmWave cellular networks can be approximated to be noise-limited rather than being interference-limited especially if the number of tiers is small. We have also shown that increasing main lobe gain results in higher SNR coverage. Moreover, we have observed the effect of biasing. Increase in the biasing factor of smaller cells has led to better coverage probability of smaller cells because of the higher number of UEs connected to them, while the overall network coverage probability has slightly diminished due to association with the BS not offering the strongest average received power. Furthermore, we have shown that smaller cells provide higher rate than larger cells. Additionally, it is verified that there is an optimal biasing factor to achieve the maximum energy efficiency. The effect of alignment error on coverage probability is also quantified. Finally, we have demonstrated that the proposed analytical framework is also applicable to $\mu$Wave-mmWave hybrid networks, and gleaned interesting insight on the impact of interference when operating in $\mu$Wave frequency bands. Investigating the effect of using different cell association techniques (e.g., which take into account the interference in a hybrid scenario) remains as future work.

\appendix
\subsection{Proof of Lemma 1} Intensity function for the $D$-ball path loss model can be computed as
\label{Proof of Lemma 1}
\begin{align}
&\Lambda_k([0,x))\stackrel{(a)}{=}\int_{\mathbb{R}^2} \mathbb{P}(L_k(r)<x)dr=2\pi\lambda_k \int_0^{\infty}\mathbb{P}((\kappa(r)r)^{\alpha^k(r)}<x)r dr \nonumber \\
&\stackrel{(b)}{=}2\pi\lambda_k\bigg(\beta_{k1} \int_0^{R_{k1}} r\mathds{1} (\kappa_1^Lr^{\alpha_1^{k,L}}<x) dr +(1-\beta_{k1})\int_0^{R_{k1}} r\mathds{1} (\kappa_1^Nr^{\alpha_1^{k,D}}<x) dr \nonumber \\
&+\beta_{k2}\int_{R_{k1}}^{R_{k2}} r\mathds{1} (\kappa_2^Lr^{\alpha_2^{k,L}}<x) dr +(1-\beta_{k2})\int_{R_{k1}}^{R_{k2}} r\mathds{1} (\kappa_2^Nr^{\alpha_2^{k,N}}<x) dr \bigg)+ \cdots \nonumber \\
&+ \beta_{kD}\int_{R_{k(D-1)}}^{R_{kD}} r\mathds{1} (\kappa_D^Lr^{\alpha_D^{k,L}}<x) dr+ (1-\beta_{kD})\int_{R_{k(D-1)}}^{R_{kD}} r\mathds{1} (\kappa_D^Nr^{\alpha_D^{k,N}}<x) dr \bigg)  \nonumber \\
&\stackrel{(c)}{=}2\pi\lambda_k \sum_{d=1}^{D}\bigg( \beta_{kd}\int_{R_{k(d-1)}}^{\min\{R_{kd},(x/\kappa_d^L)^{\frac{1}{\alpha_d^{k,L}}}\}}  rdr +(1-\beta_{kd})\int_{R_{k(d-1)}}^{\min\{R_{kd},(x/\kappa_d^N)^{\frac{1}{\alpha_d^{k,N}}}\}}  rdr \bigg) \nonumber \\
&=\pi\lambda_k \sum_{d=1}^{D} \bigg(\beta_{kd}((R_{kd}^2-R_{k(d-1)}^2)\mathds{1}(x>\kappa_d^L R_{kd}^{\alpha_d^{k,L}})+((x/\kappa_d^L)^{\frac{2}{\alpha_d^{k,L}}}-R_{k(nd-1)}^2) \nonumber \\
&\mathds{1}(\kappa_d^L R_{k(d-1)}^{\alpha_d^{k,L}}<x<\kappa_d^L R_{kd}^{\alpha_d^{k,L}}))+(1-\beta_{kd})((R_{kd}^2-R_{k(d-1)}^2)\mathds{1}(x>\kappa_d^N R_{kd}^{\alpha_d^{k,N}}) \nonumber \\
&+((x/\kappa_d^N)^{\frac{2}{\alpha_d^{k,N}}}-R_{k(d-1)}^2)\mathds{1}(\kappa_d^N R_{k(d-1)}^{\alpha_d^{k,N}}<x<\kappa_d^N R_{kd}^{\alpha_d^{k,N}}))\bigg).
\end{align}
where (a) follows from the definition of intensity function for the point process of the path loss $\mathcal{N}_k=\{L_k(r)\}_{r \in \phi_k}$; (b) is obtained when different values of distance dependent path loss exponent $\alpha_k(r)$ are inserted according to the $D$-ball model; and (c) follows from the definition of the indicator function. Finally, evaluating the integrals and rearranging the terms, we obtain the result in Lemma 1.

\subsection{Proof of Lemma 3}
\label{Proof of Lemma 3}
Note that the association probability is
\begin{align}
& \mathcal{A}_{k,s}=\mathbb{P}(P_k G_k B_k L_{k,s}^{-1} \geq \max_{j,j\neq k}P_j G_j B_j L_{j}^{-1}) \mathbb{P} (L_{k,s^{\prime}}>L_{k,s}) \nonumber \\
&\stackrel{(a)}{=}\bigg (\prod_{j=1,j \neq k}^K \mathbb{P}(P_k G_k B_k L_{k,s}^{-1} \geq P_j G_j B_j L_{j}^{-1})\bigg ) \mathbb{P} (L_{k,s^{\prime}}>L_{k,s}) \nonumber \\
&=\int_0^\infty \prod_{j=1,j \neq k}^K \bar{F}_{L_j}(\frac{P_j B_j}{P_k G_k B_k} l_{k,s}) e^{-\Lambda_{k,s^{\prime}}([0,l_{k,s}))}f_{L_{k,s}}(l_{k,s})dl_{k,s} \nonumber \\
&\stackrel{(b)}{=}\int_0^\infty e^{-\sum_{j=1,j \neq k}^K \Lambda_j([0,\frac{P_j B_j}{P_k G_k B_k} l_{k,s}))} e^{-\Lambda_{k,s^{\prime}}([0,l_{k,s}))} \Lambda_{k,s}^{\prime}([0,l_{k,s})) e^{-\Lambda_{k,s}([0,l_{k,s}))}dl_{k,s} \nonumber \\
&\stackrel{(c)}{=} \hspace{-0.1cm} \int_0^\infty e^{-\sum_{j=1,j \neq k}^K \Lambda_j([0,\frac{P_j G_j B_j}{P_k G_k B_k} l_{k,s}))} \Lambda_{k,s}^{\prime}([0,l_{k,s}))e^{-\Lambda_{k}([0,l_{k,s}))}dl_{k,s} \nonumber \\
&=\int_0^\infty \hspace{-0.2cm} \Lambda_{k,s}^{\prime}([0,l_{k,s})) e^{-\sum_{j=1}^K \Lambda_j([0,\frac{P_j G_j B_j}{P_k G_k B_k} l_{k,s}))} dl_{k,s},
\end{align}
where $s, s' \in \{\LOS,\NLOS\}$, and $s \neq s'$. In (a), CCDF of $L_j$ is formulated as a result of the first probability expression, and similarly $\mathbb{P} (L_{k,s^{\prime}}>L_{k,s})=\bar{F}_{L_{k,s^{\prime}}}(l_{k,s})=e^{-\Lambda_{k,s^{\prime}}([0,l_{k,s}))}$; (b) follows from the definition of the CCDF of the path loss, and by plugging the PDF of the path loss  $L_{k,s}$; and (c) follows from the fact that $\Lambda_{k,s}([0,l_{k,s}))+\Lambda_{k,s^{\prime}}([0,l_{k,s}))=\Lambda_{k}([0,l_{k,s}))$.

\subsection{Proof of Theorem 1}
\label{Proof of Theorem 1}
The coverage probability can be expressed as
\begin{align}
\PC&=\sum_{k=1}^K \sum_{s \in {\LOS,\NLOS}}\bigg[\mathbb{P}(\SINR_{k,s}>\Gamma_k;t=k)\mathbb{P}(L_{k,s^{\prime}}>L_{k,s})\bigg], \nonumber \\
&=\sum_{k=1}^K \sum_{s \in {\LOS,\NLOS}}\bigg[\underbrace{\mathbb{P}(\SINR_{k,s}>\Gamma_k)}_{\PC^{k,s}(\Gamma_k)}  \underbrace{\mathbb{P}(P_k G_k B_k L_{k,s}^{-1} \geq \max_{j,j\neq k}P_j G_j B_j L_{j}^{-1}) \mathbb{P}(L_{k,s^{\prime}}>L_{k,s})}_{\mathcal{A}_{k,s}}\bigg], \nonumber \\ \label{1}
\end{align}
where the last step follows from the assumption that $\Phi_j$ and $\Phi_k$ are independent from each other for $j\neq k$. The expression to obtain the association probability, $\mathcal{A}_{k,s}$ was provided in Lemma 3. Given that the UE is associated with a BS in $\Phi_{k,s}$, the conditional coverage probability $\PC^{k,s}(\Gamma_k)$ can be computed as follows
\begin{align}
&\PC^{k,s}(\Gamma_k)=\mathbb{P}(\SINR_{k,s}>\Gamma_k) \nonumber \\
&\hspace{-0.0cm}=\mathbb{P}\bigg (\frac{P_kG_0h_{k,0}L_{k,s}^{-1}}{\sigma_k^2+\sum_{j=1}^K \sum_{i \in \Phi_{j}\setminus{B_{k,0}}} P_j G_{j,i} h_{j,i} L_{j,i}^{-1}(r)} >\Gamma_k \bigg) \nonumber \\
&\hspace{-0.0cm}=\mathbb{P}\bigg (h_{k,0}>\frac{\Gamma_k L_{k,s}}{P_kG_0}\bigg(\sigma_k^2+\sum_{j=1}^K \bigg(I_{j,\LOS}+I_{j,\NLOS}\bigg)\bigg)\bigg) \nonumber \\
&\hspace{-0.0cm} \approx  \sum_{n=1}^{N_s} (-1)^{n+1} {N_s \choose n} e^{-u\sigma_k^2} \prod_{j=1}^K \bigg( \mathcal{L}_{I_{j,\LOS}}(u)\mathcal{L}_{I_{j,\NLOS}}(u)\bigg),  \label{2}
\end{align}
where $u=\frac{n\eta_s\Gamma_k L_{k,s}}{P_kG_0}$, $I_{j,s}=\sum_{i \in \Phi_{j,s}\setminus{B_{k,0}}} P_j G_{j,i} h_{j,i} L_{j,i}^{-1}(r)$ is the interference from the $j$th tier LOS and NLOS BSs, and $\mathcal{L}_{I_{j,s}}(u)$ is the Laplace transform of $I_{j,s}$ evaluated at $u$. The approximation in the last step is obtained using the same approach as in \cite[Equation (22) Appendix C]{Bai2}. Tools from stochastic geometry can be applied to compute the Laplace transform $\mathcal{L}_{I_{j,s}}(u)$ for $s \in \{\LOS,\NLOS\}$. Using the thinning property, we can split $I_{j,s}$ into three independent PPPs as follows \cite{Bai3}:
\begin{equation}
I_{j,s}=I_{j,s}^{MM}+I_{j,s}^{Mm}+I_{j,s}^{mm}=\sum_{G \in \{MM,Mm,mm\}} I_{j,s}^G \label{interference_spliting}
\end{equation}
where $I_{j,s}^G$ for $s \in \{\LOS,\NLOS\}$ denotes the interference from BSs with random antenna gain $G$ defined in (\ref{eq:antennagains}). According to the thinning theorem, each independent PPP has a density of $\lambda_j p_G$ where $p_G$ is given in (\ref{eq:antennagains}) for each antenna gain $G \in \{MM,Mm,mm\}$. Inserting (\ref{interference_spliting}) into the Laplace transform expression and using the definition of Laplace transform yield
\begin{align}
&\mathcal{L}_{I_{j,s}}(u)= \mathbb{E}_{I_{j,s}}[e^{-uI_{j,s}}]=\mathbb{E}_{I_{j,s}}\big[e^{-u(I_{j,s}^{MM}+I_{j,s}^{Mm}+I_{j,s}^{mm})}\big] = \prod_G \mathbb{E}_{I_{j,s}^G}[ e^{-uI_{j,s}^G}], \label{eq:LT}
\end{align}
where $G \in \{MM, Mm, mm\}$, $u=\frac{n\eta_s\Gamma_k L_{k,s}}{P_kG_0}$, and the last step follows from the fact that $I_{j,s}^G$s are the interferences generated from independent thinned PPPs. Laplace transforms of the interferences from the LOS and NLOS interfering BSs with a generic antenna gain $G$ can be calculated using stochastic geometry as follows:
\begin{align}
\mathbb{E}_{I_{j,s}^G}[ e^{-uI_{j,s}^G}]&= e^{-\int_{\frac{P_jB_j}{P_kB_k}l_{k,s}}^{\infty}(1-\mathbb{E}_{h,s} [ e^{-u P_j G h_{j,s} x^{-1} }])\Lambda_{j,s}^{\prime}(dx) } \nonumber \\
&\hspace{-1cm}\stackrel{(a)}{=} e^{- \int_{\frac{P_jB_j}{P_kB_k}l_{k,s}}^{\infty} (1-1/(1+uP_j Gx^{-1}/N_s)^{N_s})\Lambda_{j,s}^{\prime}(dx)}, \label{eq:LT_1}
\end{align}
where $\Lambda_{j,s}^{\prime}(dx)$ is obtained by differentiating the equations in (\ref{intensity_function_LOS}) and (\ref{intensity_function_NLOS}) with respect to $x$ for $s \in \{\LOS,\NLOS\}$, respectively, (a) is obtained by computing the moment generating function (MGF) of the gamma random variable $h$, and the lower bound for the integral is determined using the fact that the minimum separation between the UE and the interfering BS from the $j$th tier is equal to $\frac{P_jG_jB_j}{P_kG_kB_k}l_{k,s}$. Finally, by combining (\ref{Association_Prob}), (\ref{1}), (\ref{2}), (\ref{eq:LT}) and (\ref{eq:LT_1}), SINR coverage probability expression given in Theorem 1 is obtained.

\end{spacing}

\vspace{-.2cm}

\begin{spacing}{1.5}

\end{spacing}
\end{document}